\documentclass[prd,aps,showpacs,notitlepage,nofootinbib,superscriptaddress]{revtex4-1}
\usepackage{epsfig}
\newcommand{\ben}{\begin{eqnarray}}
\newcommand{\een}{\end{eqnarray}}
\newcommand{\nnu}{\nonumber\\}
\newcommand{\bef}{\begin{figure}[htb]\centering}
\newcommand{\eef}{\end{figure}}

\begin{document}
\title{Transverse momentum imbalance \\  of back-to-back particle production in  
p+A and e+A collisions}

\date{\today}

\author{Hongxi Xing}
\email{xinghx@iopp.ccnu.edu.cn} 
\affiliation{Institute of Particle Physics, 
                   Central China Normal University,
                   Wuhan 430079, China}

\author{Zhong-Bo Kang}
\email{zkang@lanl.gov} 
\affiliation{Theoretical Division, 
                   Los Alamos National Laboratory, 
                   Los Alamos, NM 87545, USA}

\author{Ivan Vitev}
\email{ivitev@lanl.gov}                   
\affiliation{Theoretical Division, 
                   Los Alamos National Laboratory, 
                   Los Alamos, NM 87545, USA}                         

\author{Enke Wang}
\email{wangek@iopp.ccnu.edu.cn} 
\affiliation{Institute of Particle Physics, 
                   Central China Normal University,
                   Wuhan 430079, China}

\begin{abstract}
We study the nuclear enhancement of the transverse momentum imbalance for back-to-back  
particle production in both p+A and e+A collisions. Specifically,  
we present results for photon+jet and photon+hadron production in p+A collisions, 
di-jet and di-hadron production in e+A collisions, and heavy-quark and heavy-meson pair 
production in both p+A and e+A collisions. We evaluate the effect of both initial-state 
and final-state multiple scattering, which determine the strength of the nuclear-induced 
transverse momentum imbalance in these processes. We give theoretical predictions for 
the experimentally relevant kinematic regions in d+Au collisions at RHIC, p+Pb collisions 
at LHC and e+A collisions at the future EIC and LHeC.
\end{abstract}


\maketitle

\section{Introduction}
Ultra-relativistic  nucleus-nucleus (A+A)  collisions
at the Relativistic Heavy Ion Collider (RHIC)  and the Large Hadron Collider (LHC)  
have paved the way to studying important properties of a new state-of-matter created in 
such  collisions, the quark-gluon plasma (QGP)~\cite{Gyulassy:2003mc}. In A+A reactions 
both final-state QGP effects and initial-state cold nuclear matter effects modify 
experimental  observables relative to the naive binary collision-scaled 
proton-proton (p+p) baseline expectation. To disentangle these effects has become
a top priority for the heavy ion program. Proton-nucleus  (p+A) collisions 
and electron-nucleus  (e+A) reactions are, at  present,  the only  tools that provide an 
opportunity to probe experimentally and understand theoretically 
cold nuclear matter effects without the complication of final-state interactions in the 
QGP~\cite{Boer:2011fh,pA,Kang:2011ni}.

Transverse momentum broadening in p+A and e+A collisions is among the most studied  
cold nuclear matter effects. When a fast parton propagates through  nuclear matter, it 
can accumulate additional transverse momentum via multiple scattering with the soft partons inside 
the big nucleus either before or after the hard collision. This phenomenon is known as transverse 
momentum broadening, and it can be probed experimentally through the nuclear modification of 
single inclusive jet or hadron production in e+A collisions~\cite{Guo:1998rd, Airapetian:2009jy, Domdey:2008aq}, 
the nuclear broadening of Drell-Yan di-lepton or $W^\pm/Z^0$ boson production in p+A 
collisions~\cite{Fries:2002mu,Kang:2008us} and the Cronin  effect~\cite{Accardi:2002ik,Vitev:2003xu}. 
One can also study the multiple scattering effects 
through two-particle correlations. For example, the nuclear 
enhancement of the transverse momentum imbalance of di-jet and di-hadron production 
was shown to be sensitive to both initial-state and final-state multiple scattering \cite{Kang:2011bp}.

Different theoretical approaches have been employed to compute and describe the nuclear 
broadening effect. These include the dipole approach~\cite{Dolejsi:1993iw, Johnson:2000dm}, 
the random walk approach~\cite{Baier:1996sk}, the diagrammatic Glauber multiple 
scattering~\cite{Gyulassy:2002yv}, the color glass condensate approach~\cite{Dumitru:2001jn,Albacete:2010pg}, 
soft collinear  effective theory~\cite{Idilbi:2008vm, Ovanesyan:2011xy, D'Eramo:2010ak}, and the high-twist  
power expansion approach~\cite{Luo:1993ui, Guo:1998rd, Kang:2008us, Kang:2011bp}. Some possible 
connections and relations among different frameworks have been discussed in~\cite{Raufeisen:2003zk}. Following 
our previous study \cite{Kang:2011bp},  we will evaluate the nuclear broadening in the formalism that 
represents multiple scattering as contributions to the cross section from higher twist
matrix elements in the nuclear state. This framework follows a well-established QCD factorization formalism 
for particle production in p+A collisions \cite{Qiu:1990xxa, Luo:1993ui, Luo:1992fz, Qiu:2001hj}, and
has been previously used to describe 
cold nuclear matter effects, such as energy loss~\cite{Xing:2011fb, Wang:2001ifa, Guo:2000nz},
dynamical  shadowing~\cite{Qiu:2003vd, Qiu:2004da, Kang:2007nz} and broadening 
effects~\cite{Luo:1993ui, Guo:1998rd, Kang:2008us, Kang:2011bp}. 
The purpose of our paper is to apply the techniques developed in our previous study 
of di-jet and di-hadron transverse momentum imbalance~\cite{Kang:2011bp} to 
new channels, which will be accessible to the experiments in the near future. It differs 
from more generic parton broadening phenomenology in that the color and kinematic structures 
of the hard part are evaluated exactly.  In particular, we will study the nuclear 
enhancement of the transverse momentum imbalance for 
photon+jet (photon+hadron) in p+A, di-jet (di-hadron) in e+A, and 
heavy-quark (heavy-meson) pair production in both p+A and e+A collisions. 
These two-particle correlation observables can be studied in 
d+Au collisions at $\sqrt{s}=200$~GeV at RHIC, the forthcoming p+Pb run at  
$\sqrt{s}=5$~TeV at the LHC, and at the planned  Electron Ion Collider (EIC) and Large Hadron Electron 
Collider (LHeC). Our paper presents a unified formalism to predict theoretically  this observable  
for  multiple final-state channels.

The rest of our manuscript  is organized as follows: in Sec. II we study  photon+jet 
(photon+hadron) and heavy-quark (heavy-meson) pair production in p+A collisions.  
The beginning of this section is used to introduce the basic definition of 
the transverse  momentum imbalance, which is common to all studied processes. 
We then  take into account both initial-state and final-state 
multiple scattering to compute  the nuclear enhancement in the transverse momentum 
imbalance of the produced back-to-back 
particle pair. In Sec. III we study the nuclear broadening of the transverse momentum 
imbalance for di-jet (di-hadron), and heavy-quark (heavy-meson) pair production in e+A 
collisions. In Sec. IV we present our numerical estimate on the nuclear broadening 
for the relevant kinematics at RHIC, LHC and the future EIC and LHeC. We summarize our results 
in Sec. V.

\section{Nuclear enhancement of the transverse momentum imbalance   in $p+A$ collisions}

The main purpose of this paper is to study the nuclear enhancement of the transverse momentum 
imbalance for  back-to-back  particle production in both p+A and e+A collisions, 
$h(P')+A(P)\to h_1(p_1)+h_2(p_2)+X$. Here $h$ and $A$ are the incoming hadron (or virtual 
photon) and the nucleus, respectively. $h_1$ and $h_2$ are the produced particles in the final 
state with transverse momenta $\vec{p}_{1\perp}$ and $\vec{p}_{2\perp}$. To lowest order in 
perturbative QCD, the production of these two particles arises from hard $2\to 2$ scattering 
processes. 
If we denote by $z$ the $h$ and $A$ collision axis, $h_1$ and $h_2$ are produced approximately 
back-to-back in the transverse $(x,y)$ plane: 
$\vec{p}_{1\perp}\approx -\vec{p}_{2\perp}$. In p+A collisions, however, the incoming parton 
can undergo multiple scattering before the hard collisions. The produced  final-state particles 
are also likely to undergo multiple interactions in the big nucleus if they are 
strongly interacting. Both  initial-state  and final-state multiple scattering lead to 
acoplanarity, or momentum imbalance of the observed two particles. To quantify this effect, we define 
the transverse momentum imbalance $\vec{q}_{\perp}$  as:
\ben
\vec{q}_{\perp}=\vec{p}_{1\perp}+\vec{p}_{2\perp},
\een
and the average transverse momentum squared imbalance 
\ben
\langle q_\perp^2\rangle = \left. \left(  \int d^2 \vec{q}_\perp q_\perp^2
\frac{d\sigma}{d\mathcal{PS} \, d^2\vec{q}_\perp}  \right) \right/ \frac{d\sigma}{d\mathcal{PS}}.
\label{avg-qt}
\een
Here, $d\sigma/d\mathcal{PS}$ is the differential cross section with $d\mathcal{PS}$ 
representing the relevant phase space, to be defined for each process in the 
corresponding section. For example, in p+A collisions, $d\mathcal{PS}=dy_1dy_2 dp_\perp^2$ 
for  photon+jet production and $d\mathcal{PS}=dy_1dy_2 dp_{1\perp} dp_{2\perp}$ for 
photon+hadron production.

The enhancement  of the transverse momentum imbalance (or nuclear broadening)  in 
h+A (h=p, $\gamma^*$) collisions relative to h+p collisions can be quantified by the 
difference:
\ben
\Delta \langle q_\perp^2\rangle = \langle q_\perp^2\rangle_{hA} 
- \langle q_\perp^2\rangle_{hp}.
\label{master}
\een
The broadening $\Delta \langle q_\perp^2\rangle$ is a result of multiple quark and gluon scattering, 
and is a direct probe of the nuclear medium properties. We now  take into account both 
initial-state and final-state multiple parton interactions to calculate the nuclear broadening 
$\Delta \langle q_\perp^2\rangle$ for photon+jet (photon+hadron) and heavy-quark (heavy-meson) 
pair production in p+A collisions. In the next section we study di-jet (di-hadron) and 
heavy-quark (heavy-meson) pair production in e+A collisions.


\subsection{Photon+jet (hadron) production in p+A collisions}
\subsubsection{Photon+jet production}
Consider the following back-to-back photon+jet production process (the photons here and 
throughout the paper are direct photons) in p+A collisions:
\ben
p(P') + A(P) \to \gamma(p_1)+J(p_2)+X.
\een
Here, $P',~P$ are the four momentum of the incoming hadron and nucleus (per nucleon) with 
atomic number $A$ and $p_1$ and $p_2$ are the four momentum of the produced final-state 
photon and jet, respectively. The light-cone components of the final-state particles are 
given by
\ben
p_1=\left[\frac{|\vec{p}_{1\perp}|}{\sqrt{2}}e^{y_1},~\frac{|\vec{p}_{1\perp}|}{\sqrt{2}}e^{-y_1},
~\vec{p}_{1\perp}\right],~~~~
p_2=\left[\frac{|\vec{p}_{2\perp}|}{\sqrt{2}}e^{y_2},~\frac{|\vec{p}_{2\perp}|}{\sqrt{2}}e^{-y_2},
~\vec{p}_{2\perp}\right],
\een
where $y_{1,2}$ and $\vec{p}_{1,2\perp}$ are the rapidities and transverse momenta, respectively. 
In leading-order collinear factorized perturbative QCD, the photon and the jet are produced 
exactly back-to-back, $\vec{p}_{1\perp}=-\vec{p}_{2\perp}$. It is important to realize that to 
this order the jet is idetical to the leading  parton. 
It is only at next-to-leading order (NLO) that the QCD structure of the jet starts to play 
a role in the theoretical description of physics observables. In heavy ion collisions, 
experimental observables that include jets in the final state are presented after subtraction
of the uncorrelated soft hadronic background. Thus, they would be directly comparable to the
results presented in this paper. The differential cross section to leading order
can be written as \cite{Owens:1986mp}
\ben
\frac{d\sigma}{dy_1 dy_2 dp^2_\perp}
=\frac{\pi\alpha_s\alpha_{em}}{s^2}\sum_{a,b}
\frac{f_{a/p}(x') f_{b/A}(x)}{x' x} H^{U}_{ab\to\gamma d}(\hat s, \hat t, \hat u) \, ,
\label{Eq:main}
\een
where $\sum_{a, b}$ runs over all possible parton flavors, $s=(P'+P)^2$ is the 
center-of-mass energy squared, $f_{a/p}$ and $f_{b/A}$ represent the proton and 
nuclear parton distribution functions, respectively. At this order in perturbation theory,  
the parton momentum  fractions $x'$ and $x$ are uniquely related  to the rapidities 
and the jet transverse momentum:
\ben
x'=\frac{p_\perp}{\sqrt{s}}\left(e^{y_1}+e^{y_2}\right), 
\qquad
x=\frac{p_\perp}{\sqrt{s}}\left(e^{-y_1}+e^{-y_2}\right).
\een
\bef
\psfig{file=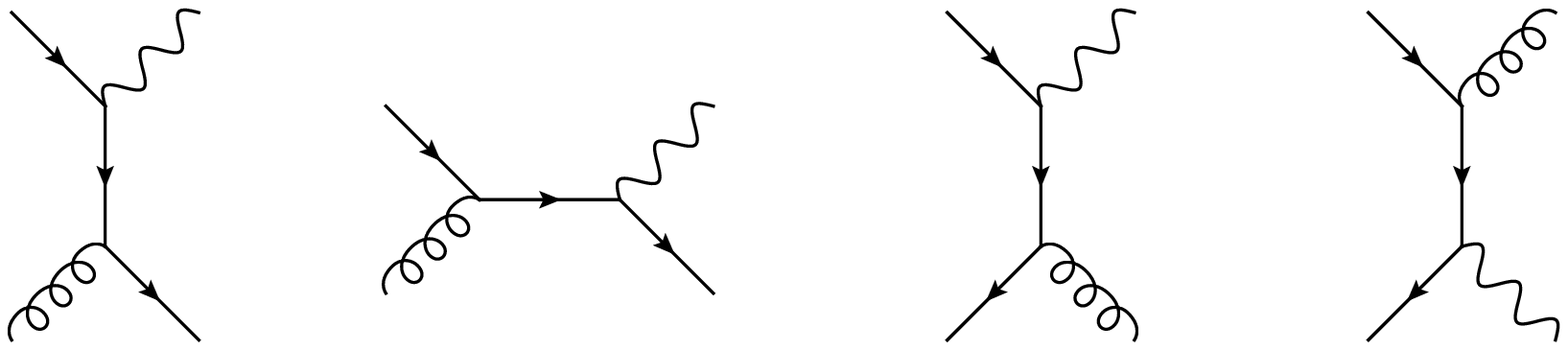, width=4in}
\caption{Leading order Feynman diagrams for photon-jet production.}
\label{fig:single}
\eef
$H^{U}_{ab\to cd}(\hat s, \hat t, \hat u)$ are the partonic cross sections as a function of 
the usual partonic Mandelstam variables $\hat s, \hat t, \hat u$. They are calculated from 
the Feynman diagrams in Fig.~\ref{fig:single} and are given by \cite{Owens:1986mp,Kang:2011rt}
\ben
H^U_{qg\to \gamma q}&=&e_q^2 \frac{1}{N_c}\left[-\frac{{\hat s}}{\hat t}
-\frac{\hat t}{{\hat s}}\right],
\\
H^U_{gq\to \gamma q}&=&e_q^2 \frac{1}{N_c}\left[-\frac{{\hat s}}{\hat u}
-\frac{\hat u}{{\hat s}}\right],
\\
H^U_{q\bar q\to \gamma g}&=&e_q^2\frac{N_c^2-1}{N_c^2}\left[\frac{\hat t}{\hat u}
+\frac{\hat u}{\hat t}\right],
\een
where $N_c=3$ is the number of colors.
\bef
\psfig{file=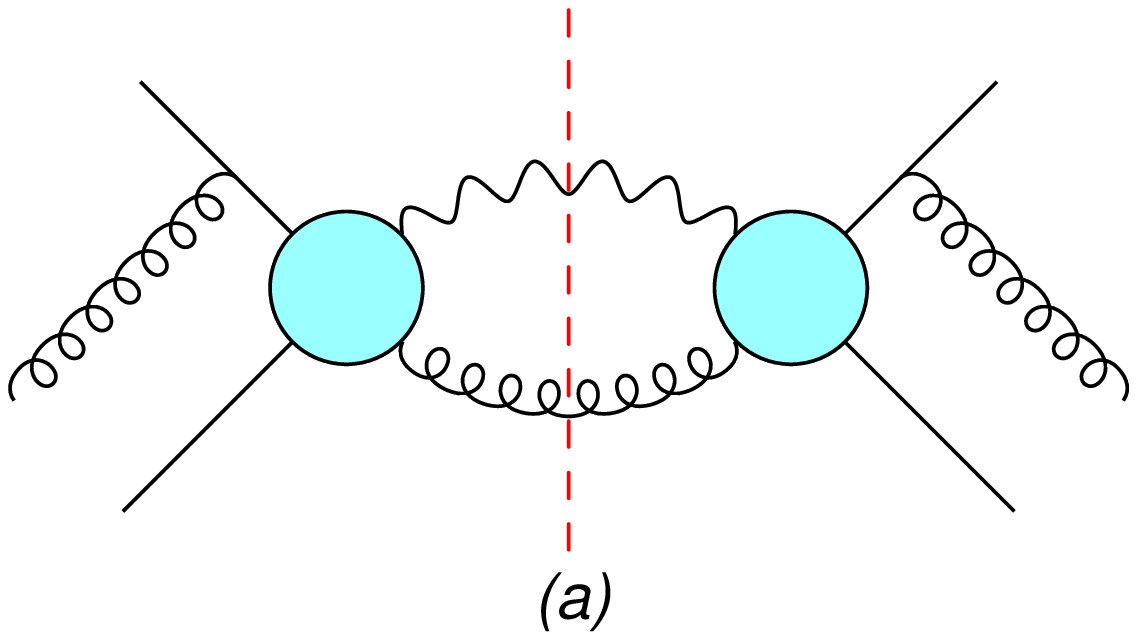, width=1.75in}
\hskip 0.15in
\psfig{file=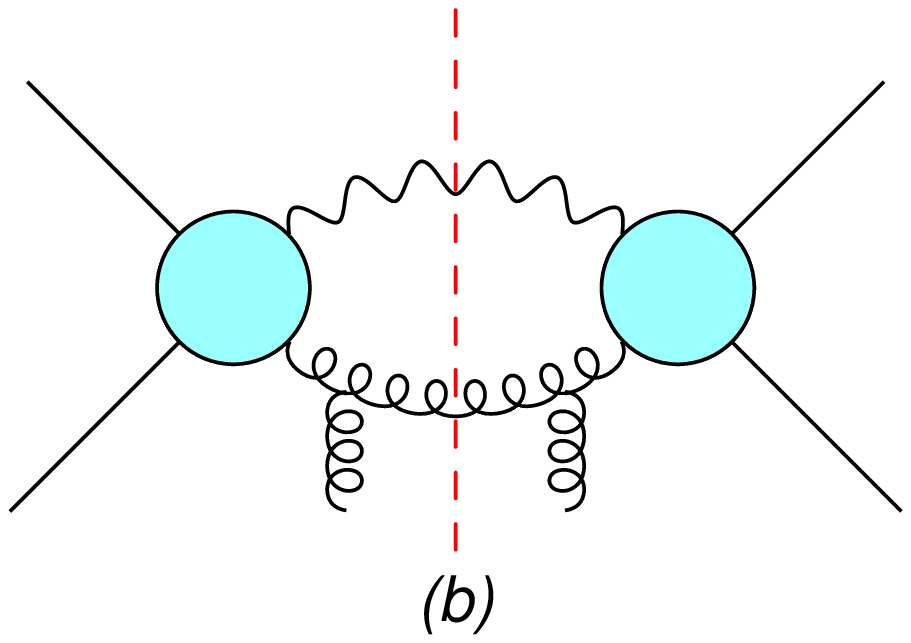, width=1.4in}
\hskip 0.15in
\psfig{file=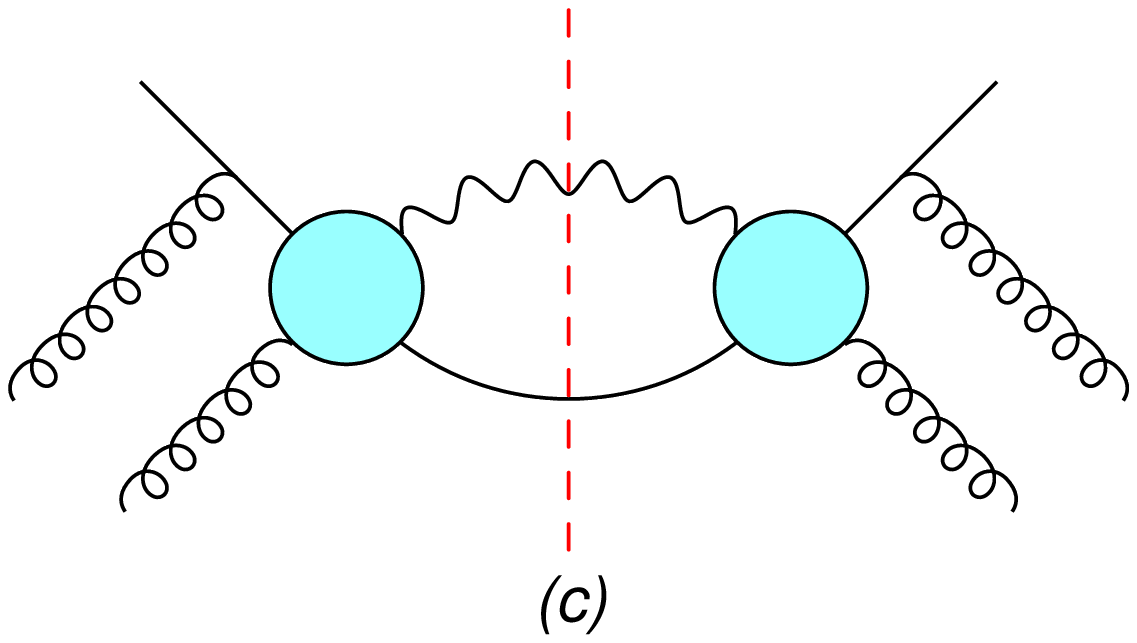, width=1.75in}
\hskip 0.15in
\psfig{file=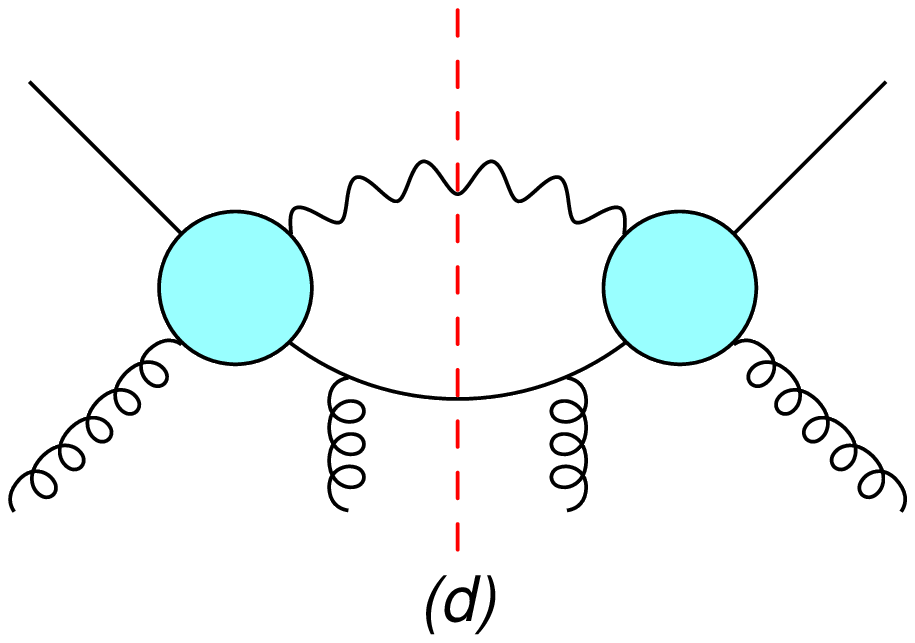, width=1.4in}
\caption{Double scattering diagrams for $q{\bar q}\to \gamma g$ (left two) and $qg\to \gamma q$ 
(right two): (a) and (c) are for initial-state double scattering, while (b) and (d) are for 
final-state double scattering. The blobs represent the tree-level diagrams as shown in 
Fig.~\ref{fig:single}.}
\label{fig:pAphotonjet}
\eef

In p+A collisions, the energetic incoming parton from the proton can undergo multiple scattering 
with the soft partons inside the nuclear matter before the hard collisions  
(initial-state multiple scattering). After the hard collisions, the leading outgoing parton 
(opposite the photon) will also  undergo multiple interactions in the large nucleus 
(final-state multiple scattering). These interactions  lead to an enhancement in the 
photon+jet transverse momentum imbalance, which can be quantified by $\Delta \langle q_\perp^2\rangle$, 
as defined in Eq.~(\ref{master}). This nuclear broadening $\Delta \langle q_\perp^2\rangle$ 
can be calculated in perturbative QCD. A specific method based on double parton scattering  
has been discussed in detail in 
Refs.~\cite{Luo:1993ui, Guo:1998rd, Kang:2008us, Kang:2011bp}. Our derivation closely 
follows our previous paper \cite{Kang:2011bp}. The leading contribution to the nuclear 
broadening comes from the double scattering: either initial-state double scattering, as in 
Fig.~\ref{fig:pAphotonjet} (a) and (c)
for the partonic channels $q{\bar q}\to \gamma g$ and $qg\to \gamma q$, respectively; 
or final-state double scattering,  as in Fig.~\ref{fig:pAphotonjet} (b) and (d). 
We calculate the contributions from these diagrams in the covariant gauge and obtain the following 
expression for the nuclear broadening of photon+jet production in p+A collisions:
\ben
\Delta\langle q_\perp^2\rangle
=\left(\frac{8\pi^2\alpha_s}{N_c^2-1}\right)
\frac{ \sum_{a, b}\frac{f_{a/p}(x')}{x' x}
\left[T_{b/A}^{(I)}(x) H^I_{ab\to\gamma d} (\hat s, \hat t, \hat u)+
T_{b/A}^{(F)}(x) H^F_{ab\to\gamma d} (\hat s, \hat t, \hat u)\right]}
{\sum_{a,b}\frac{f_{a/p}(x') f_{b/A}(x)}{x' x} 
H^{U}_{ab\to\gamma d}(\hat s, \hat t, \hat u)},
\label{imbalance}
\een
where $T_{b/A}^{(I)}(x) = T_{q/A}^{(I)}(x)$ (or $T_{g/A}^{(I)}(x)$) are twist-4 quark-gluon (or gluon-gluon) correlation 
functions associated with initial-state multiple scattering, with the following operator 
definitions~\cite{Luo:1993ui, Guo:1998rd, Kang:2008us, Kang:2011bp}:
\ben
T_{q/A}^{(I)}(x) &=&
 \int \frac{dy^{-}}{2\pi}\, e^{ixp^{+}y^{-}}
 \int \frac{dy_1^{-}dy_{2}^{-}}{2\pi} \,
      \theta(y^{-}-y_1^{-})\,\theta(-y_{2}^{-})
     \frac{1}{2}\,
     \langle p_{A}|F_{\alpha}^{\ +}(y_{2}^{-})\bar{\psi}_{q}(0)
                  \gamma^{+}\psi_{q}(y^{-})F^{+\alpha}(y_1^{-})
     |p_{A} \rangle,
\label{TqA}
\\
T_{g/A}^{(I)}(x) &=&
 \int \frac{dy^{-}}{2\pi}\, e^{ixp^{+}y^{-}}
 \int \frac{dy_1^{-}dy_{2}^{-}}{2\pi} \,
      \theta(y^{-}-y_1^{-})\,\theta(-y_{2}^{-})
\frac{1}{xp^+}\,
\langle p_A| F_\alpha^{~+}(y_2^-)
F^{\sigma+}(0)F^+_{~\sigma}(y^-)F^{+\alpha}(y_1^-)|p_A\rangle\, .
\label{TgA}
\een
On the other hand, $T_{q/A}^{(F)}(x)$ and $T_{g/A}^{(F)}(x)$ are the corresponding twist-4 
correlation functions connected with 
the final-state multiple scattering. They are given by the same expressions in 
Eqs.~(\ref{TqA}) and (\ref{TgA}), except for 
the $\theta$-functions that are replaced as follows \cite{Kang:2008us, Kang:2011bp}:
\ben
\theta(y^{-}-y_1^{-})\,\theta(-y_{2}^{-})
\to
\theta(y_1^{-}-y^{-})\,\theta(y_{2}^{-}).
\label{theta}
\een
The hard part functions $H^I_{ab\to\gamma d} $ and $H^F_{ab\to\gamma d}$ are associated with initial and final-state multiple 
scattering, respectively; and are given by
\ben
\label{HI}
H^I_{ab\to\gamma d} &=& \left\{
  \begin{array}{l l}
    C_F H^U_{ab\to\gamma d} & \quad \text{a=quark}\\
     \\
    C_A H^U_{ab\to\gamma d} & \quad \text{a=gluon}\\
  \end{array} \right.  \; , 
  \\
  \nnu
  \nnu
H^F_{ab\to\gamma d} &=& \left\{
  \begin{array}{l l}
    C_F H^U_{ab\to\gamma d} & \quad \text{d=quark}\\
     \\
    C_A H^U_{ab\to\gamma d} & \quad \text{d=gluon}\\
  \end{array} \right.   \; .
 \label{photon-final}
\een
Here, $C_F=(N_c^2-1)/2N_c$ and $C_A=N_c$ are the quadratic Casimir in the fundamental and adjoint 
representations of SU(3)$_c$, respectively. It is also instructive to recall that the strength of 
the multiple scattering depends on the color representation of the complete scattered parton
system~\cite{Kang:2011bp}. For photon+jet production, since the photon does not carry color, the 
final-state multiple scattering only depends on the color of the jet parton (whether it is a quark 
or a gluon), as can be clearly seen in Eq.~(\ref{photon-final}).

\subsubsection{Photon+hadron production}
For the back-to-back photon+hadron production in p+A collisions, $p(P') + A(P) \to \gamma(p_1)+h(p_2)+X$, 
the leading order  differential cross section has the following form:
\ben
\frac{d\sigma}{dy_1 dy_2 dp_{1\perp} dp_{2\perp}}=\frac{2\pi\alpha_s\alpha_{em}}{s^2}\sum_{abd}D_{h/d}(z)
\frac{f_{a/p}(x') f_{b/A}(x)}{x' x} H^{U}_{ab\to\gamma d}(\hat s, \hat t, \hat u),
\label{dihadron}
\een
where the momentum fractions $z$, $x'$, and $x$ are given by
\ben
z=\frac{p_{2\perp}}{p_{1\perp}},
\qquad
x'=\frac{p_{1\perp}}{\sqrt{s}}\left(e^{y_1}+e^{y_2}\right), 
\qquad
x=\frac{p_{1\perp}}{\sqrt{s}}\left(e^{-y_1}+e^{-y_2}\right).
\een
Following our previous paper \cite{Kang:2011bp}, we can easily generalize the calculation for nuclear broadening 
$\Delta\langle q_\perp^2\rangle$ in photon+jet to photon+hadron production by including the fragmentation function $D_{h/d}(z)$, 
and it is given by
\ben
\Delta\langle q_\perp^2 \rangle =\left(\frac{8\pi^2\alpha_s}{N_c^2-1}\right)
\frac{ \sum_{abd}D_{h/d}(z) \frac{f_{a/p}(x')}{x' x}
\left[T_{b/A}^{(I)}(x) H^I_{ab\to\gamma d} (\hat s, \hat t, \hat u)+
T_{b/A}^{(F)}(x) H^F_{ab\to\gamma d} (\hat s, \hat t, \hat u)\right]}
{\sum_{abd} D_{h/d}(z)\frac{f_{a/p}(x') f_{b/A}(x)}{x' x} H^{U}_{ab\to\gamma d}(\hat s, \hat t, \hat u)}. 
\een

\subsection{Heavy-quark (heavy-meson) pair production in p+A collision}
\subsubsection{Heavy-quark pair production}
We now study the heavy-quark pair production, $p(P')+A(P)\to Q(p_1)+\bar{Q}(p_2)+X$. At leading order in perturbative QCD, 
the heavy quark $Q$ and anti-quark $\bar{Q}$ are produced back-to-back
through the following partonic channels: $q\bar{q}\to Q\bar{Q}$,  $gg\to Q\bar{Q}$. 
Thus, $\vec{p}_{1\perp}=-\vec{p}_{2\perp}$ and $|\vec{p}_{1\perp}|=|\vec{p}_{2\perp}|\equiv p_\perp$. 
The differential cross section can be written as
\ben
\frac{d\sigma}{dy_1 dy_2 dp^2_\perp}
=\frac{\pi\alpha_s^2}{s^2}\sum_{a,b}
\frac{f_{a/p}(x') f_{b/A}(x)}{x' x} H^{U}_{ab\to Q\bar{Q}}(\hat s, \hat t, \hat u) \, ,
\een
where the parton momentum fractions $x'$ and $x$ are given by
\ben
x'=\frac{m_\perp}{\sqrt{s}}\left(e^{y_1}+e^{y_2}\right), 
\qquad
x=\frac{m_\perp}{\sqrt{s}}\left(e^{-y_1}+e^{-y_2}\right),
\een
with $m_\perp=\sqrt{p_\perp^2+m_Q^2}$ and $m_Q$ the heavy quark mass. The hard part functions $H^{U}_{ab\to {Q}d}$,  
$H^{U}_{ab\to \bar{Q}d}$,  $H^{U}_{ab\to Q\bar{Q}}$ for the 
partonic processes  $a(p_a)+b(p_b) \to Q(p_1)+d(p_2)$,  $a(p_a)+b(p_b)\to \bar{Q}(p_1)+d(p_2)$,   
 $a(p_a)+b(p_b)\to Q(p_1)+\bar{Q}(p_2)$  
relevant to the variable flavor scheme  \cite{Olness:1997yc}  are given in \cite{Vitev:2006bi}. In this paper
we work in the fixed flavor scheme (three light quarks). Furthermore,  if instead of the 
standard definition of Mandelstam variables one introduces the notation
\ben
\hat{s}=(p_a+p_b)^2,
\qquad
\hat{t}=(p_a-p_1)^2-m_Q^2,
\qquad
\hat{u}=(p_b-p_1)^2-m_Q^2, 
\een
the two relevant hard part functions can be written compactly as \cite{Kang:2008ih}
\ben
H^U_{q\bar{q}\to Q\bar{Q}}&=&\frac{C_F}{N_c}\left[\frac{\hat{t}^2+\hat{u}^2+2m_Q^2\hat{s}}{\hat{s}^2}\right],\nnu
H^U_{gg\to Q\bar{Q}}&=&\frac{1}{2N_c}\left[\frac{1}{\hat{t}\hat{u}}-\frac{N_c}{C_F}\frac{1}{\hat{s}^2}\right]
\left[\hat{t}^2+\hat{u}^2+4m_Q^2\hat{s}-\frac{4m_Q^4\hat{s}^2}{\hat{t}\hat{u}}\right].
\een

The nuclear enhancement  of the transverse momentum imbalance $\Delta\langle q_\perp^2\rangle$ in p+A collisions 
can be easily calculated and the final result is given by
\ben
\Delta\langle q_\perp^2\rangle
=\left(\frac{8\pi^2\alpha_s}{N_c^2-1}\right)
\frac{ \sum_{a, b}\frac{f_{a/p}(x')}{x' x}
\left[T_{b/A}^{(I)}(x) H^I_{ab\to Q\bar{Q}} (\hat s, \hat t, \hat u)+
T_{b/A}^{(F)}(x) H^F_{ab\to Q\bar{Q}} (\hat s, \hat t, \hat u)\right]}
{\sum_{a,b}\frac{f_{a/p}(x') f_{b/A}(x)}{x' x} 
H^{U}_{ab\to Q\bar{Q}}(\hat s, \hat t, \hat u)},
\een
where the hard functions $H^I_{ab\to Q\bar{Q}}$ and $H^F_{ab\to Q\bar{Q}}$ are again associated with initial-state 
and final-state multiple scattering and are given by
\ben
H^I_{q\bar{q}\to Q\bar{Q}} &=& C_F H^U_{q\bar{q}\to Q\bar{Q}},
\\
H^I_{gg\to Q\bar{Q}} &=& C_A H^U_{gg\to Q\bar{Q}},
\\
H^F_{q\bar{q}\to Q\bar{Q}} &=& C_A H^U_{q\bar{q}\to Q\bar{Q}},
\\
H^F_{gg\to Q\bar{Q}} &=& C_A H^U_{gg\to Q\bar{Q}} - \frac{1}{2(N_c^2-1)}\frac{1}{\hat{t}\hat{u}}
\left[\hat{t}^2+\hat{u}^2+4m_Q^2\hat{s}-\frac{4m_Q^4\hat{s}^2}{\hat{t}\hat{u}}\right].
\een
When the heavy quark mass $m_Q\to 0$, we recover the published results for $H^{I, F}_{q\bar{q}\to q'\bar{q}'}$ 
and $H^{I,F}_{gg\to q\bar{q}}$ in our previous paper \cite{Kang:2011bp}.

\subsubsection{Heavy-meson pair production}
One can easily extend the above calculation to heavy-meson pair production, such as back-to-back $D$+$\bar{D}$, 
$p(P')+A(P)\to D(p_1)+\bar{D}(p_2)+X$. The differential cross section is given by \cite{Vitev:2006bi}
\ben
\frac{d\sigma}{dy_1 dy_2 dp_{1\perp} dp_{2\perp}}=\frac{2\pi\alpha_s^2}{s^2}\sum_{a,b}
\int \frac{dz_1}{z_1}D_{D/Q}(z_1) D_{\bar{D}/\bar{Q}}(z_2) 
\frac{f_{a/p}(x') f_{b/A}(x)}{x' x} H^{U}_{ab\to Q\bar{Q}}(\hat s, \hat t, \hat u),
\een
where $D_{D/Q}(z_1)$ and $D_{\bar{D}/\bar{Q}}(z_2) $ are heavy-meson fragmentation functions, 
 $z_2=z_1\, p_{2\perp}/p_{1\perp}$, and:
\ben
x'=\frac{m_{\perp}}{\sqrt{s}}\left(e^{y_1}+e^{y_2}\right), \qquad 
x=\frac{m_{\perp}}{\sqrt{s}}\left(e^{-y_1}+e^{-y_2}\right),
\een
with $m_\perp=\sqrt{(p_{1\perp}/z_1)^2+m_Q^2}$. The nuclear broadening $\Delta\langle q_\perp^2\rangle$ is given by
\ben
\Delta\langle q_\perp^2 \rangle =\left(\frac{8\pi^2\alpha_s}{N_c^2-1}\right)
\frac{ \sum_{a,b}\int \frac{dz_1}{z_1}D_{D/Q}(z_1) D_{\bar{D}/\bar{Q}}(z_2)  \frac{f_{a/p}(x')}{x' x} 
\left[T_{b/A}^{(I)}(x) H^I_{ab\to Q\bar{Q}} (\hat s, \hat t, \hat u)+
T_{b/A}^{(F)}(x) H^F_{ab\to Q\bar{Q}} (\hat s, \hat t, \hat u)\right]}
{\sum_{a,b}\int \frac{dz_1}{z_1}D_{D/Q}(z_1) D_{\bar{D}/\bar{Q}}(z_2) 
\frac{f_{a/p}(x') f_{b/A}(x)}{x' x} H^{U}_{ab\to Q\bar{Q}}(\hat s, \hat t, \hat u)}.\qquad
\een

Including the di-jet and di-hadron production in our previous paper \cite{Kang:2011bp}, we have 
computed the nuclear  broadening in the transverse momentum imbalance for all important 
back-to-back two-particle production channels in p+A collisions. 
We now study two-particle production in e+A collisions in the next section.

\section{Nuclear enhancement of the transverse momentum imbalance  in $e+A$ collisions}
In this section, we use the same approach to calculate the nuclear broadening  for di-jet and 
di-hadron, as well as heavy-quark (heavy-meson) pair production in deep inelastic scattering (DIS) 
of a lepton on a big nucleus, or  $\gamma^*+A$ collisions. Since the virtual photon 
does not interact with the soft partons in the nuclear target 
via the  strong force, the nuclear broadening only comes from the final-state multiple scattering.

\subsection{Di-jet (di-hadron) production in DIS}
\subsubsection{Di-jet production}
Nuclear enhancement of the di-jet transverse momentum imbalance in photo-production has been 
calculated in Ref.~\cite{Luo:1993ui}. 
Here we will generalize result to di-jet production in DIS,
\ben
\gamma^*(P_{\gamma^*}) + A(P) \to J_1(p_1) + J_2(p_2) +X,
\een
where the incoming virtual-photon $\gamma^*$ carries momentum $P_{\gamma^*}$ with 
virtuality $P_{\gamma^*}^2=-Q^2$.  $P$ is 
the four momentum of the target nucleus, $p_1$ and $p_2$ are the momenta of 
final-state jets $J_1$ and $J_2$, respectively. 
We will work in the center of mass frame of $\gamma^*$+A, in which the light-cone 
components of the incoming particles are 
\ben
P_{\gamma^*} = \left[\sqrt{\frac{s}{2}}, -\frac{Q^2}{\sqrt{2s}}, 0_\perp\right],
\qquad
P= \left[0, \frac{s+Q^2}{\sqrt{2s}}, 0_\perp\right].
\een
At leading order, the two jets have opposite transverse momentum but with 
the same magnitude $p_\perp$, and the  differential cross section can be written as
\ben
\frac{d\sigma}{dy_1 dp^2_{\perp}}=\frac{\pi\alpha_s\alpha_{em}}{\left(s+Q^2\right)^2}\sum_{b}
\frac{1}{1-\frac{p_{\perp}}{\sqrt{s}}e^{y_1}}\frac{f_{b/A}(x)}{x} H^{U}_{\gamma^* b\to cd}(\hat s, \hat t, \hat u, Q^2)\, ,
\label{singlephoto}
\een
where the momentum fraction $x$ is given by
\ben
x = x_B + \frac{p_\perp \sqrt{s}}{s+Q^2} \left[e^{-y_1}+ \frac{1}{\sqrt{s}/p_\perp - e^{y_1}}\right] ,
\een
with $x_B=Q^2/2P\cdot P_{\gamma^*}=Q^2/(s+Q^2)$. $y_1$ is the rapidity of the first jet $J_1$, and the rapidity $y_2$ of 
the second jet $J_2$ is related to $y_1$ as follows:
\ben
y_2 = \ln\left(\frac{\sqrt{s}}{p_\perp} - e^{y_1}\right).
\een
The hard function $H^{U}_{\gamma^* b\to cd}$ is given by
\ben
H^U_{\gamma^* q\to qg}&=&e_q^2\frac{N_c^2-1}{N_c} \left[-\frac{{\hat s}}{\hat t}-\frac{\hat t}{{\hat s}}+\frac{2\hat u Q^2}{{\hat s}{\hat t}}\right],
\\
H^U_{\gamma^* g\to q\bar q}&=&e_q^2\left[\frac{\hat t}{\hat u}+\frac{\hat u}{\hat t}
-\frac{2\hat s Q^2}{{\hat t}{\hat u}}\right],
\een
where $\hat s$, $\hat t$, and $\hat u$ is defined as
\ben
\hat s=(P_{\gamma^*}+x\,P)^2,
\qquad
\hat t=(P_{\gamma^*}-p_1)^2,
\qquad
\hat u=(x\, P-p_1)^2.
\een
Since the virtual photon does not interact strongly  with the nucleus,  
the broadening $\Delta\langle q_\perp^2\rangle$ 
is only sensitive to  final-state multiple scattering and is given by
\ben
\Delta\langle q_\perp^2\rangle=\left(\frac{8\pi^2\alpha_s}{N_c^2-1}\right)
\frac{\sum_{b}\frac{1}{x} T_{b/A}^{(F)}(x) H^F_{\gamma^* b\to cd} (\hat s, \hat t, \hat u, Q^2)}
{\sum_{b}\frac{1}{x}f_{b/A}(x) H^{U}_{\gamma^* b\to cd}(\hat s, \hat t, \hat u, Q^2)}.
\een
The hard function $H^F_{\gamma^*b\to cd}$ can be written as
\ben
\label{HFphoto}
H^F_{\gamma^*b\to cd} = \left\{
  \begin{array}{l l}
    C_F H^U_{\gamma^* b\to cd} & \quad \text{b=quark}\\
     \\
    C_A H^U_{\gamma^* b\to cd} & \quad \text{b=gluon}\\
  \end{array} \right. \; .
\een
These expressions suggest that even though $H^F_{\gamma^*b\to cd}$ are the hard functions 
associated with final-state multiple 
scattering, the strength of the broadening depends on the color representation of the 
initial-state parton $b$: the color factor $C_F$ 
(or $C_A$) corresponds to the incoming quark (or gluon). This is not surprising since the rescattering effects 
are only sensitive to the total color of the final two-parton composite state, which is equal to 
the color of the initial parton $b$ (as $\gamma^*$ carries no color).
It is easy to show that by setting $Q^2\to 0$, we recover the result for di-jet photo-production derived in 
Ref.~\cite{Luo:1993ui}.

\subsubsection{Di-hadron production}
For back-to-back hadron pair production, $\gamma^*(P_{\gamma^*}) + A(P) \to h_1(p_1) + h_2(p_2) +X$, the differential cross section can be written as
\ben
\frac{d\sigma}{dy_1 dy_2dp_{1\perp} dp_{2\perp}}=\frac{2\pi\alpha_s\alpha_{em}}
{\left(s+Q^2\right)^2}\sum_{b} D_{h_1/c}(z_1) D_{h_2/d}(z_2)
\frac{f_{b/A}(x)}{x} H^{U}_{\gamma^* b\to cd}(\hat s, \hat t, \hat u, Q^2)\, ,
\een
where the momentum fractions $z_1$, $z_2$, and $x$ are given by
\ben
z_1 = \frac{p_{1\perp}}{\sqrt{s}}\left(e^{y_1}+e^{y_2}\right), 
\qquad
z_2 = \frac{p_{2\perp}}{\sqrt{s}}\left(e^{y_1}+e^{y_2}\right), 
\qquad
x = x_B+ \frac{s}{s+Q^2}\frac{e^{-y_1}+e^{-y_2}}{e^{y_1}+e^{y_2}}.
\een
The nuclear broadening $\Delta\langle q_\perp^2\rangle$ has the following form:
\ben
\Delta\langle q_\perp^2\rangle=\left(\frac{8\pi^2\alpha_s}{N_c^2-1}\right)
\frac{\sum_{b} D_{h_1/c}(z_1) D_{h_2/d}(z_2) 
\frac{1}{x} T_{b/A}^{(F)}(x) H^F_{\gamma^* b\to cd} (\hat s, \hat t, \hat u, Q^2)}{
\sum_{b} D_{h_1/c}(z_1) D_{h_2/d}(z_2)
\frac{1}{x}f_{b/A}(x) H^{U}_{\gamma^* b\to cd}(\hat s, \hat t, \hat u, Q^2)}.
\een

\subsection{Heavy-quark (heavy-meson) pair production in DIS}
\subsubsection{Heavy-quark pair production}
We now  study the heavy-quark pair production, $\gamma^*(P_{\gamma^*}) + A(P) \to Q(p_1) + \bar{Q}(p_2) +X$. 
At leading order, there is only one partonic channel that contributes, $\gamma^*+g\to Q+\bar Q$ \cite{Kang:2008qh}. 
The differential cross section can be written as
\ben
\frac{d\sigma}{dy_1 dp^2_{\perp}}=\frac{\pi\alpha_s\alpha_{em}}{\left(s+Q^2\right)^2}
\frac{1}{1-\frac{m_{\perp}}{\sqrt{s}}e^{y_1}}\frac{f_{g/A}(x)}{x} H^{U}_{\gamma^* g\to Q\bar{Q}}(\hat s, \hat t, \hat u, Q^2, m_Q^2)\, ,
\een
where $y_1$ is the rapidity of the heavy quark $Q$, while the heavy anti-quark $\bar{Q}$ has rapidity 
$y_2 = \ln\left(\sqrt{s}/{m_\perp} - e^{y_1}\right)$. The momentum fraction $x$ is given by 
\ben
x = x_B + \frac{m_\perp \sqrt{s}}{s+Q^2} \left[e^{-y_1}+ \frac{1}{\sqrt{s}/m_\perp - e^{y_1}}\right],
\een
with the transverse mass $m_\perp=\sqrt{p_\perp^2+m_Q^2}$. The hard part functions $H^U_{\gamma^* g\to Q\bar Q}$ has the following 
form,
\ben
H^U_{\gamma^* g\to Q\bar Q}=e_q^2\left[\frac{\hat t}{\hat u}+\frac{\hat u}{\hat t}
-\frac{2\hat{s}Q^2}{{\hat t}{\hat u}}
-\frac{2m_Q^2}{\hat t \hat u}\left(\frac{2m_Q^2({\hat t}+{\hat u})^2}{{\hat t}{\hat u}}
-Q^2\left(\frac{\hat t}{\hat u}+\frac{\hat u}{\hat t}\right)+2({\hat t}+{\hat u})\right)\right].
\een
Once again  $\hat s$, $\hat t$, and $\hat u$ differ slightly form the standard Mandelstam variables 
and are given by
\ben
\hat s=(P_{\gamma^*}+x\,P)^2,
\qquad
\hat t=(P_{\gamma^*}-p_1)^2-m_Q^2,
\qquad
\hat u=(x\, P-p_1)^2-m_Q^2.
\een
The nuclear enhancement of the transverse momentum imbalance can be written as
\ben
\Delta\langle q_\perp^2\rangle=\left(\frac{8\pi^2\alpha_s}{N_c^2-1}\right)
\frac{\frac{T_{g/A}^{(F)}(x)}{x} H^F_{\gamma^* g\to Q\bar{Q}}(\hat s, \hat t, \hat u, Q^2, m_Q^2)}
{\frac{f_{g/A}(x)}{x} H^{U}_{\gamma^* g\to Q\bar{Q}}(\hat s, \hat t, \hat u, Q^2, m_Q^2)} ,
\een
where $H^F_{\gamma^* g\to Q\bar Q} = C_A H^U_{\gamma^* g\to Q\bar Q}$ with $C_A$ reflective of the color 
representation of the $Q\bar{Q}$ system.

\subsubsection{Heavy-meson pair production}
One can also easily generalize the above calculation to heavy-meson pair production, such as 
back-to-back $D+\bar{D}$ production, $\gamma^*(P_{\gamma^*}) + A(P) \to D(p_1) + \bar{D}(p_2) +X$. 
The differential cross section is given by
\ben
\frac{d\sigma}{dy_1 dy_2dp_{1\perp} dp_{2\perp}}=\frac{2\pi\alpha_s\alpha_{em}}{\left(s+Q^2\right)^2} 
D_{D/Q}(z_1) D_{\bar{D}/\bar{Q}}(z_2)
\frac{f_{g/A}(x)}{x} H^{U}_{\gamma^* g\to Q\bar{Q}}(\hat s, \hat t, \hat u, Q^2)\, ,
\een
where the momentum fractions $z_1$ and $z_2$ in heavy-meson fragmentation functions are given by
\ben
z_1 = \frac{p_{1\perp}}{r\sqrt{s}}\left(e^{y_1}+e^{y_2}\right), 
\qquad
z_2 = \frac{p_{2\perp}}{r\sqrt{s}}\left(e^{y_1}+e^{y_2}\right).
\een
Here the parameter $r$ depends on the heavy quark mass $m_Q$:
\ben
r=\sqrt{1-\left[\frac{m_Q}{\sqrt{s}}(e^{y_1}+e^{y_2})\right]^2}.
\een
The nuclear broadening  for this final state is  given by
\ben
\Delta\langle q_\perp^2\rangle=\left(\frac{8\pi^2\alpha_s}{N_c^2-1}\right)
\frac{D_{D/Q}(z_1) D_{\bar{D}/\bar{Q}}(z_2)
\frac{T_{g/A}^{(F)}(x)}{x} H^F_{\gamma^* g\to Q\bar{Q}}(\hat s, \hat t, \hat u, Q^2)}{
D_{D/Q}(z_1) D_{\bar{D}/\bar{Q}}(z_2)
\frac{f_{g/A}(x)}{x} H^{U}_{\gamma^* g\to Q\bar{Q}}(\hat s, \hat t, \hat u, Q^2)}.
\een
We have now completed all  evaluations  of the nuclear  enhancement of the transverse momentum imbalance 
in back-to-back two-particle production in both p+A and e+A collisions. We will use these expressions to present  
predictions    for   $ \Delta\langle q_\perp^2\rangle $ relevant to  future experimental measurements  
in the next section.


\section{Numerical results}

In this section we  present phenomenological applications of our results. Specifically,
we give theoretical predictions for the nuclear broadening (or enhancement of the transverse 
momentum imbalance) for back-to-back 
particle production in d+Au collisions at RHIC, for the forthcoming p+Pb collisions at LHC, 
and for the e+A collisions at the 
planned EIC and LHeC. The only new unknown ingredient in our calculation is the twist-4 
quark-gluon and gluon-gluon correlation functions. Following~\cite{Kang:2011bp, Qiu:2003vd, Qiu:2004da}, 
we parametrize them as follows:
\ben
\frac{4\pi^2\alpha_s}{N_c}\,T_{q,g/A}^{(I)}(x)=\frac{4\pi^2\alpha_s}{N_c}\,T_{q,g/A}^{(F)}(x)
= \xi^2 \left(A^{1/3}-1\right) f_{q,g/A}(x),
\label{eq:ht}
\een
where $f_{q,g/A}(x)$ is the standard  leading-twist parton distribution function for quarks or gluons, 
respectively. In Eq.~(\ref{eq:ht}) $\xi^2=0.12$ GeV$^2$ represents a characteristic scale of parton multiple 
scattering and was extracted from deep inelastic scattering data~\cite{Qiu:2003vd}.  
The definition in Eq.~(\ref{eq:ht}) is such that $\xi^2 \left(A^{1/3}-1\right)$ 
can be thought of a dynamical quark mass generated in the background soft gluon field of the nucleus
in minimum-bias reactions~\cite{Qiu:2004qk}. Our implementation  has also been successful in  describing 
the nuclear suppression of single inclusive hadron  production~\cite{Qiu:2004da,Vitev:2006bi} and the 
di-hadron transverse momentum imbalance and correlations 
in d+Au collisions at forward rapidities at RHIC $\sqrt{s}=200$ GeV~\cite{Kang:2011bp}.

As demonstrated in Ref.~\cite{Qiu:2003vd}, by approximately decomposing a nuclear state into a product 
of nucleon states  the parameter $\xi^2$ can be expressed in terms of an averaged gluon field strength 
squared $\langle F^{+\alpha} F^+_{~~\alpha}\rangle\sim \lim_{x\to 0} x \, f_{g/A}(x)$, the 
soft-gluon number density. In this picture, 
$\xi^2$ represents the strength of the multiple scattering, thus proportional to the number of the soft gluons 
in the nuclear medium. While at tree level $\xi^2$ is a fixed number, higher order corrections may
 provide energy dependence to this parameter (for example, at LHC energies  $\xi^2$ can be larger 
in comparison to the one used  RHIC energies). We study this possibility phenomenologically by presenting
results for a range  of  $\xi^2$ from 0.12 GeV$^2$ to 0.20 GeV$^2$ at the LHC (yellow band). For the upper 
limit we take guidance from the growth of the inelastic scattering cross section, from
$\sigma_{in}^{\rm RHIC}=42$ mb at $\sqrt{s}=200$ GeV 
to $\sigma_{in}^{\rm LHC}=70$ mb at $\sqrt{s}=5$ TeV \cite{Miller:2007ri, d'Enterria:2003qs}.   
The energy dependence of $\xi^2$ (if any) can be tested and further constrained once experimental 
data at different center-of-mass energies become 
available~\footnote{Recent comparison of theoretical predictions for the cross section 
modification in p+Pb reactions \cite{Kang:2012kc} to new ALICE preliminary 
data \cite{ALICE}  
does not favor a growth of $\xi^2$. }.
To estimate the transverse momentum broadening  for the new channels derived in the last two sections, 
we  use the CTEQ6L for nucleon parton distribution functions~\cite{Pumplin:2002vw}, and EPS08 
parametrization for nuclear parton distribution functions (nPDFs)~\cite{Eskola:2008ca}. Since 
the nuclear broadening $\Delta\langle q_\perp^2\rangle$ is a ratio of cross sections as in 
Eq.~(\ref{master}), it has very weak dependence on the parametrization for nPDFs.

\bef
\psfig{file=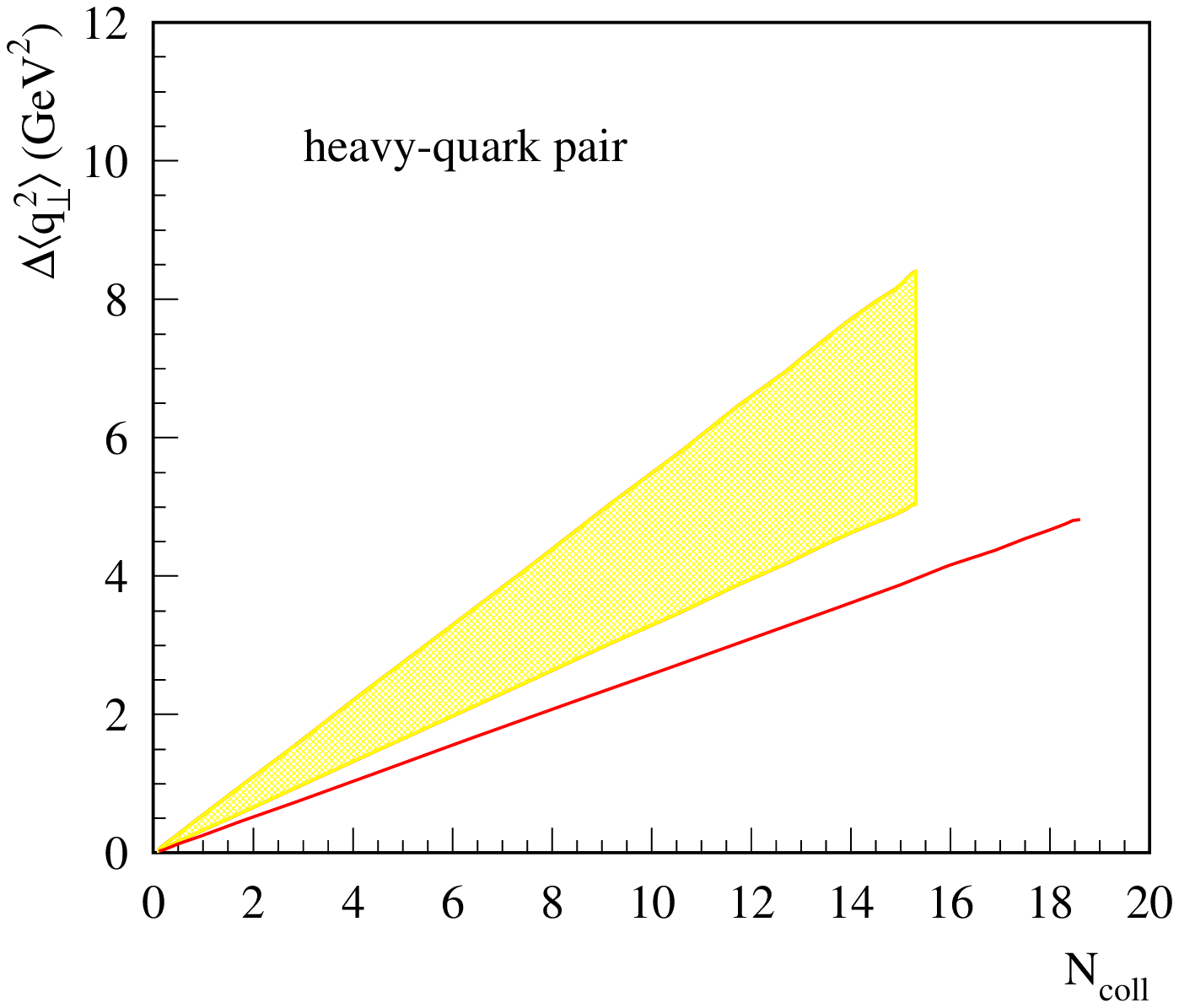, width=3in}
\hskip 0.3in
\psfig{file=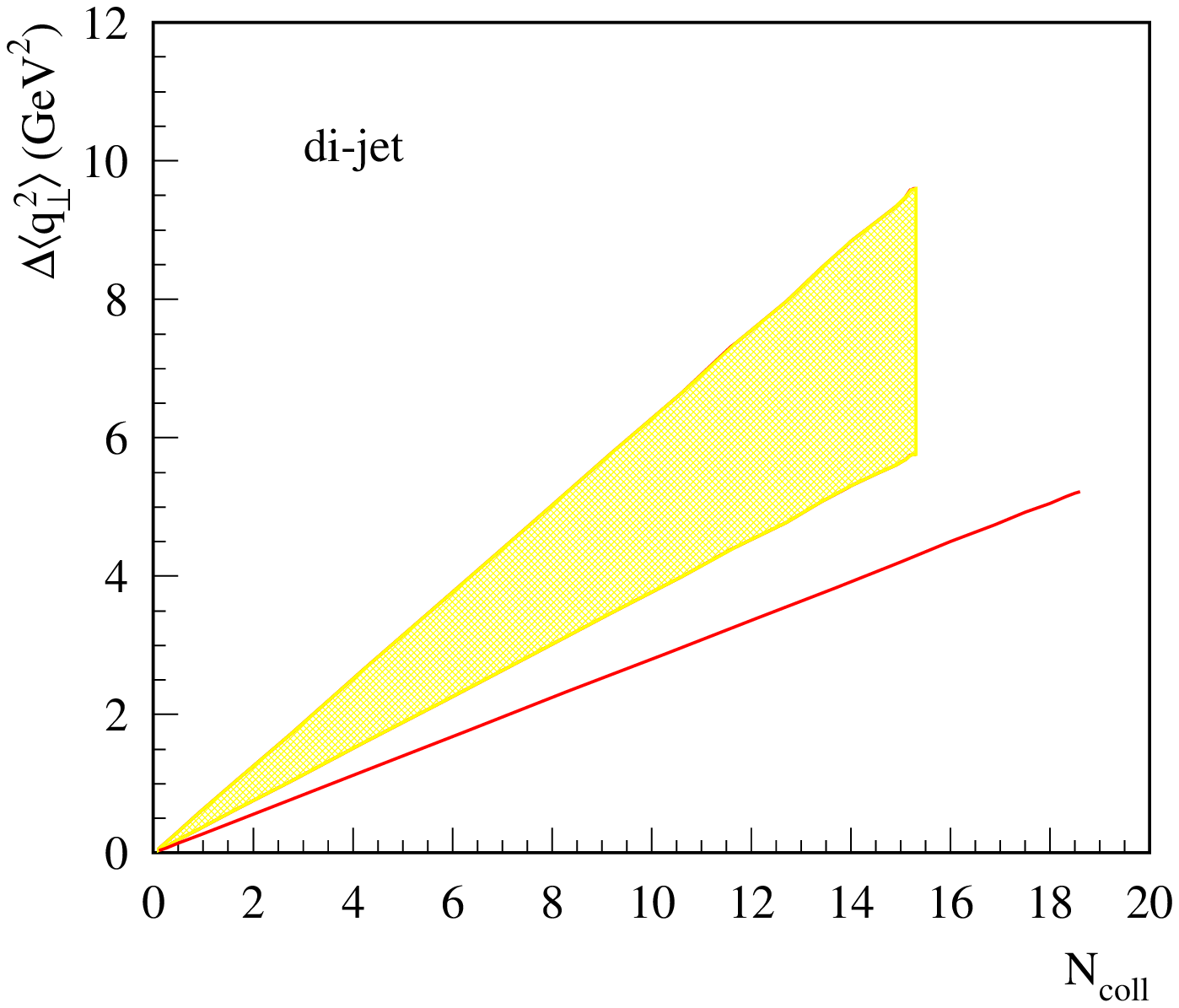, width=3in}
\caption{Nuclear broadening $\Delta\langle q_{\perp}^2\rangle$ for back-to-back heavy-quark pairs (left) and 
di-jets (right) in p+A collisions as a function of binary collision number $N_{\rm coll}$. 
We choose  rapidities $y_1=y_2=2$ for the $\sqrt{s}=5$ TeV  LHC p+Pb collisions and $y_1=y_2=1$ 
for the $\sqrt{s}=200$ GeV RHIC d+Au collisions. For the LHC, the jet transverse momentum is 
integrated over 30~GeV~$<p_\perp<$~40~GeV, 
while for RHIC the jet transverse momentum is integrated over 15~GeV~$<p_\perp<$~25~GeV. The yellow band is 
for LHC kinematics, with the band representing a variation of $\xi^2$ parameter from 0.12 GeV$^2$ to 0.20 GeV$^2$. 
The red solid curve is for RHIC kinematics with $\xi^2=0.12$ GeV$^2$.}
\label{fig:heavy}
\eef

\bef
\psfig{file=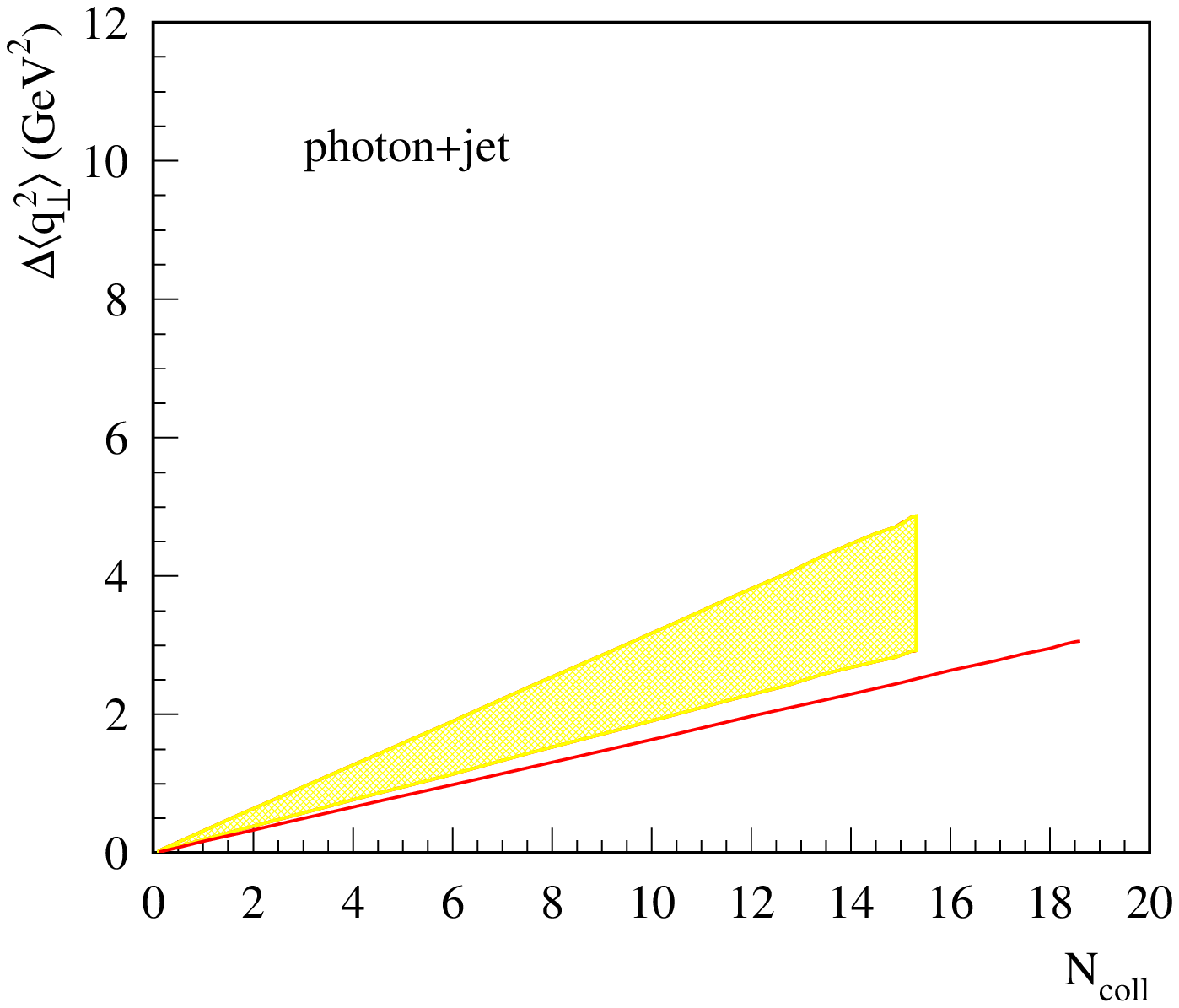, width=3in}
\hskip 0.3in
\psfig{file=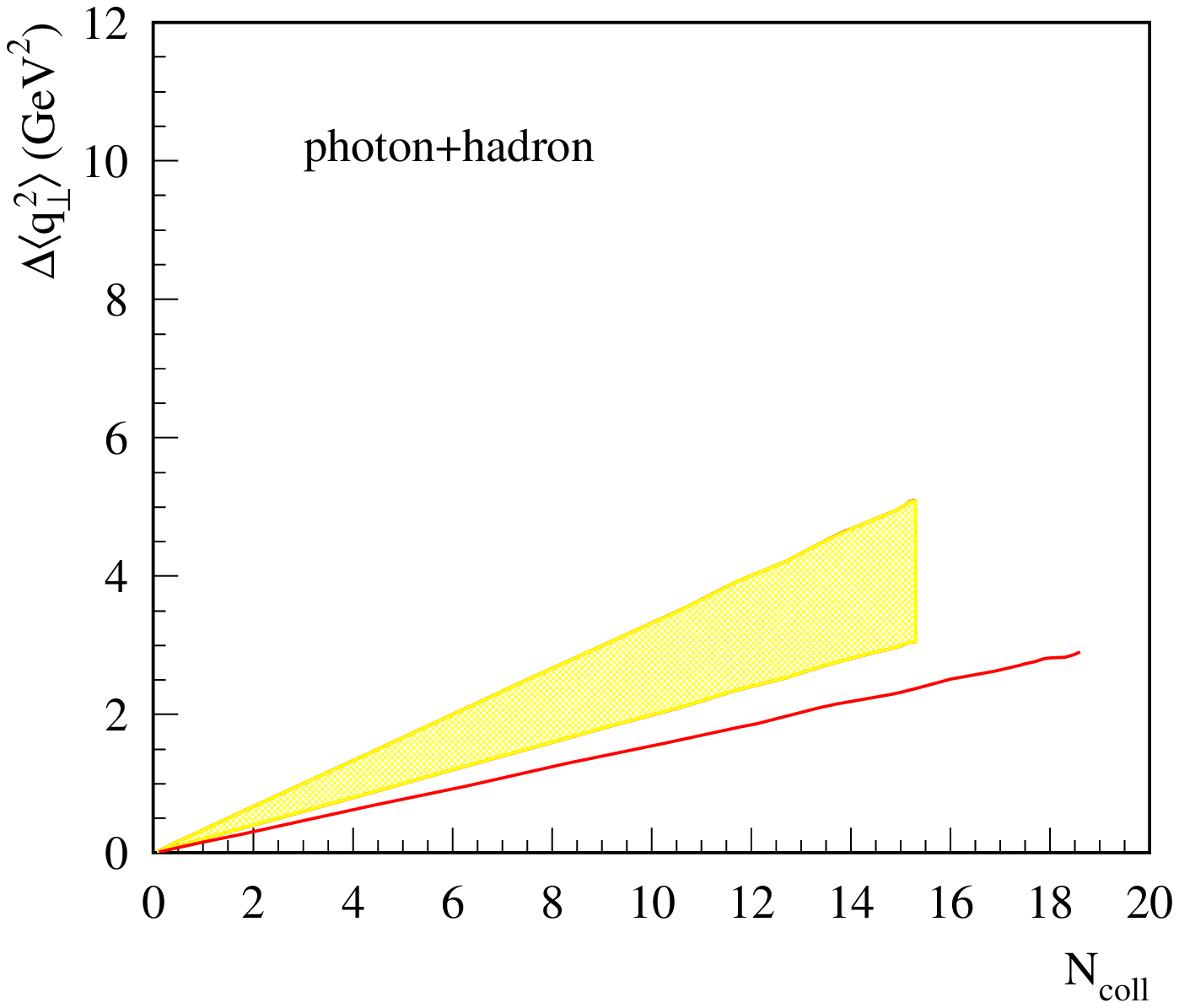, width=3in}
\caption{Left panel: nuclear broadening $\Delta\langle q_{\perp}^2\rangle$ for back-to-back photon+jet
 production in p+A collisions as a function of the binary collision number $N_{\rm coll}$. 
We choose  rapidities $y_1=y_2=2$ for the $\sqrt{s}=5$ TeV  LHC p+Pb collisions and $y_1=y_2=1$ for the $\sqrt{s}=200$~GeV RHIC d+Au collisions. For the LHC, the jet transverse momentum is integrated over 30 GeV $<p_\perp<$ 40~GeV, while for RHIC the jet transverse momentum is integrated over 15 GeV $<p_\perp<$ 25 GeV. 
Right panel: same as left plot, but now for back-to-back photon+hadron production. For LHC kinematics, 
we integrate over $10<p_{\gamma\perp}<20$ GeV and $5<p_{h\perp}<10$ GeV. For RHIC kinematics, we integrate over $5<p_{\gamma\perp}<15$ GeV and $5<p_{h\perp}<10$ GeV. 
The yellow band is 
for LHC kinematics, with the band representing a variation of $\xi^2$ parameter from 0.12~GeV$^2$ to 0.20~GeV$^2$. The red solid curve is for RHIC kinematics with $\xi^2=0.12$ GeV$^2$.}
\label{fig:photonjet}
\eef

In Fig.~\ref{fig:heavy} (left) we plot the nuclear enhancement of the the transverse 
momentum  imbalance  $\Delta\langle q_{\perp}^2\rangle$  for a  back-to-back heavy-quark pair  
in p+A collisions as a function of the binary collision number $N_{\rm coll}$. 
To take into account the centrality 
dependence, we have replaced $(A^{1/3}-1)$ by 
$(A^{1/3}-1)\langle N_{\rm coll}(b)\rangle/\langle N_{\rm coll}(b_{\rm min.bias})\rangle$. 
We choose the rapidities of both the heavy quark and the anti-quark to be $y_1=y_2=2$, and 
integrate over the transverse momentum 30~GeV $<p_\perp<$ 40~GeV for LHC p+Pb run at $\sqrt{s}=5$ TeV, 
and $y_1=y_2=1$ and 15~GeV $<p_\perp<$ 25~GeV for RHIC d+Au run at $\sqrt{s}=200$ GeV. 
The yellow band is for LHC kinematics, with the band representing a variation of the $\xi^2$ parameter from 
0.12 GeV$^2$ to 0.20 GeV$^2$ as we discussed above. The red solid curve is for 
RHIC kinematics with $\xi^2=0.12$ GeV$^2$.
For comparison, we also plot in Fig.~\ref{fig:heavy} (right) the nuclear broadening 
for di-jet production using the same kinematics. 
We find that the nuclear broadening  $\Delta\langle q_{\perp}^2\rangle$ is slightly stronger for  
di-jet production when  compared to heavy-quark pair production. This is because there are more partonic 
channels  that contribute to the di-jet. In particular, the $ gg \to gg $ channel is the most important 
and  generates the largest broadening~\cite{Kang:2011bp}.  
However, this channel does not contribute to  
heavy-quark pair production. 
Since the $gg\to gg$ channel becomes more important at larger center-of-mass energies,
we expect the difference in $\Delta\langle q_{\perp}^2\rangle$  between heavy-quark pair production 
and di-jet production to become slightly larger in going form RHIC to the LHC. 
This can be seen clearly in our predictions in Fig.~\ref{fig:heavy}.

In Fig.~\ref{fig:photonjet} (left) we plot  $\Delta\langle q_{\perp}^2\rangle$ 
for the photon+jet final state in p+A collision as a function of $N_{\rm coll}$. Compared to 
the heavy quark or di-jet production in Fig.~\ref{fig:heavy},  the nuclear broadening  is much smaller.
 This is because there is much stronger final-state multiple scattering in di-jet or heavy 
quark pair production:  both outgoing partons interact with the cold nuclear matter. 
The difference in $\Delta\langle q_{\perp}^2\rangle$   between photon+jet and di-jet 
(or heavy-quark pair) production is a direct prediction of our approach. A comparative 
experimental study facilitated by future experimental measurements will be a very useful 
test of our formalism. In Fig.~\ref{fig:photonjet} (right), we also provide a plot for 
back-to-back photon+hadron production in p+A collisions as a function of $N_{\rm coll}$. 
We find that the magnitude of the nuclear enhancement in the transverse momentum imbalance 
is very similar between photon+jet and photon+hadron production. This is expected since the 
nuclear broadening as defined in Eqs.~(\ref{avg-qt}) and (\ref{master}) is a ratio of cross 
sections and should not be affected much by the fragmentaiton function (the hadronization process).

\bef
\psfig{file=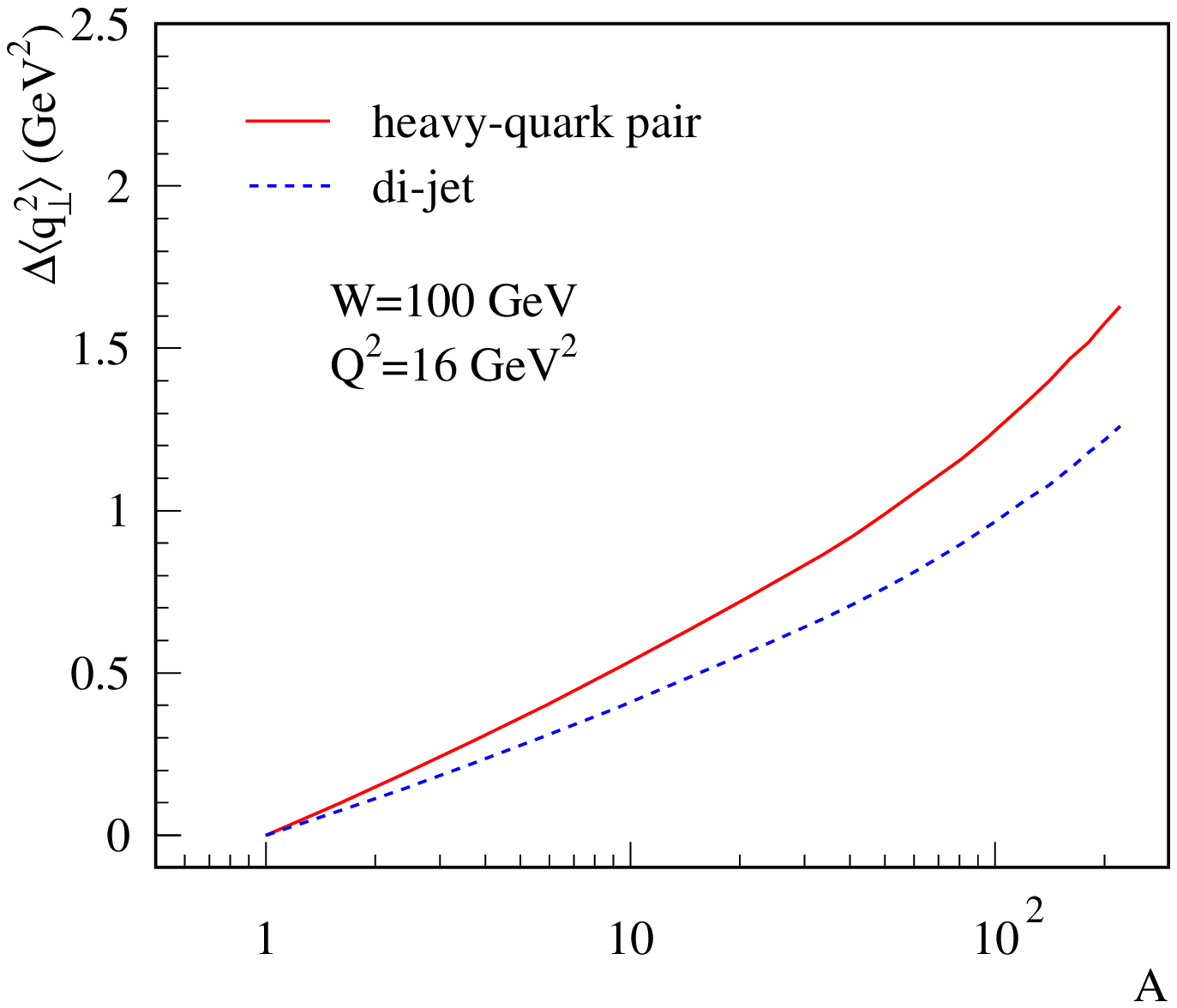, width=3in}
\hskip 0.3in
\psfig{file=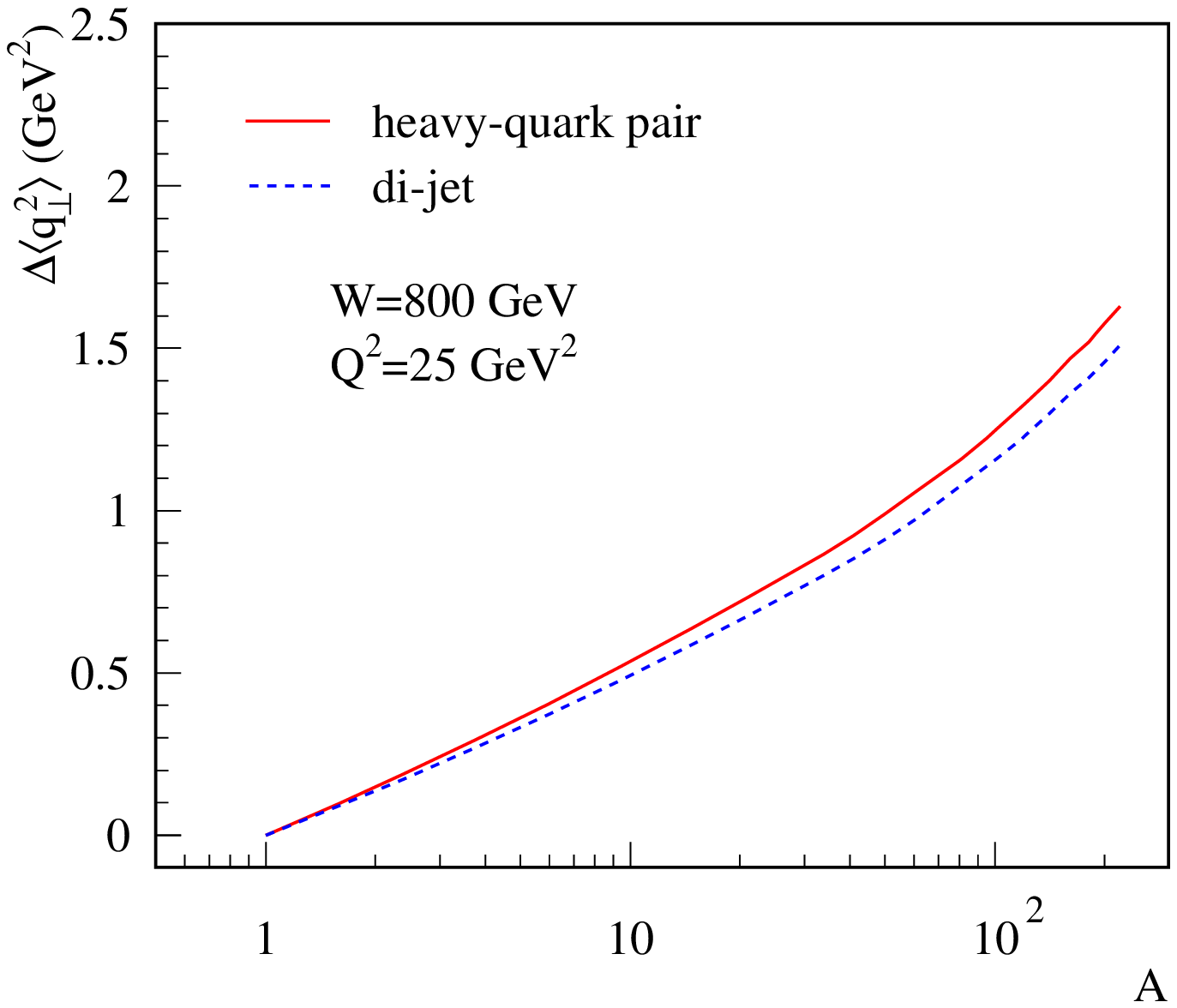, width=3in}
\caption{Transverse momentum imbalance increase $\Delta\langle q_{\perp}^2\rangle$ in $\gamma^*+A$ collision 
as a function of the atomic number $A$ at $\gamma^*$-A center-of-mass energy. 
Left panel: $\sqrt{s}=100$ GeV and photon virtuality  $Q^2$=16 GeV$^2$ for  typical EIC kinematics. 
We choose the jet rapidity $y_1=2$ and have integrated the jet transverse momentum over 
5 GeV $<p_\perp<$ 10 GeV.
Right panel:  $\sqrt{s}=800$ GeV and photon virtuality  $Q^2$=25 GeV$^2$ for  typical LHeC 
kinematics~\cite{AbelleiraFernandez:2012cc}. We choose the jet rapidity $y_1=2$ 
and have integrated the jet transverse momentum over 10 GeV $<p_\perp<$ 20 GeV.
Note that the $\gamma^*$-A center of mass energy corresponds to the standard DIS 
kinematic variable $W$: $s=(P_{\gamma^*}+P)^2=W^2$.
The solid curve is for heavy-quark pair production, while the dashed curve is for di-jet production.}
\label{fig:meson}
\eef

Nuclear-enhanced transverse momentum imbalance in di-jet and heavy-quark pair production  in DIS 
can provide a clean measurement of final-state rescattering effect. In Fig.~\ref{fig:meson} we plot 
 $\Delta\langle q_{\perp}^2\rangle$ for both di-jet and heavy-quark pair production in $\gamma^*+A$ collisions 
 as a function of the atomic number $A$. In the left panel 
we present predictions for the relevant kinematics region for the planned future EIC. 
We choose the $\gamma^*+A$ center-of-mass energy 
$\sqrt{s}$=100 GeV and the photon virtuality $Q^2=16$ GeV$^2$. Note that the $\gamma^*$-A center-of-mass 
energy corresponds to the standard DIS kinematic variable $W$: $s=(P_{\gamma^*}+P)^2=W^2$. 
We choose the jet rapidity to be $y_1=2$ and 
integrate the jet transverse momentum over 5 GeV $<p_\perp<$ 10 GeV. 
In the right panel we present predictions for the relevant kinematics region for the planned future LHeC.
We choose the $\gamma^*+A$ center-of-mass energy 
$\sqrt{s}$=800 GeV and the photon virtuality $Q^2=25$ GeV$^2$, and integrate the jet 
transverse momentum over 10 GeV $<p_\perp<$ 20 GeV.
Our numerical results show that contrary to the case in p+A collisions, the growth of the 
transverse momentum imbalance for heavy-quark pair production is larger than that for  di-jet  
production because of the different color factors of the rescattering final-state. 
In the heavy-quark case, only one process is present at leading order and the initial-state parton 
is a gluon. This gives rise to a color factor $C_A$. On the other hand, there are two processes 
that contribute to the di-jet transverse momentum imbalance and they give rise to a combination 
of color factors: $C_F$ and $C_A$. Consequently, the nuclear broadening is smaller than the one 
observed in  heavy-quark pair production.

\section{Summary}
Within a high-twist approach to parton interactions in cold nuclear matter we studied the nuclear 
enhancement of the transverse momentum imbalance for photon+jet and photon+hadron production in 
p+A collisions, di-jet and di-hadron production in e+A collisions, and heavy-quark (heavy-meson) 
pair production in both p+A and e+A collisions. By taking into account both initial-state and 
final-state multiple scattering, we derived results to lowest order in perturbative QCD for the 
increase in the transverse momentum imbalance of two particle-production for these channels. 
We presented numerical predictions for the kinematic regions relevant to d+Au collision at RHIC, 
p+Pb collisions at LHC, and e+A collisions at a future EIC and LHeC. We found that the nuclear broadening 
in photon+jet (photon+hadron) production is much smaller than the one in di-jet (or heavy-quark pair) 
production, due to weaker final-state multiple scattering. It is also interesting to notice 
that in p+A collisions the di-jet accumulates more nuclear-induced transverse momentum 
imbalance than the heavy-quark pair production, while in e+A collisions it is the other way around.
The difference in the nuclear broadening  among the various channels  is a direct predictions of 
our approach. We emphasize that a comparative study of the transverse momentum imbalance of 
back-to-back  particle production, facilitated by  future experimental measurements, 
will be a valuable probe for the multiple scattering effect in cold nuclear matter 
and a test of our theoretical formalism.

\section*{Acknowledgments}
This research is supported by the US Department of
Energy, Office of Science, under Contract No.~DE-AC52-06NA25396, and in part 
by the LDRD program at LANL, NSFC of China under Project No. 10825523.

\end{document}